# Random Access in Uplink Massive MIMO Systems: How to exploit asynchronicity and excess antennas

Luca Sanguinetti*†, Antonio A. D'Amico*, Michele Morelli*, Merouane Debbah†§

*Dipartimento di Ingegneria dell'Informazione, University of Pisa, Pisa, Italy
†Large Networks and System Group (LANEAS), CentraleSupélec, Université Paris-Saclay, Gif-sur-Yvette, France
§Mathematical and Algorithmic Sciences Lab, Huawei France, Paris, France

*Abstract*—Massive MIMO systems, where the base stations are equipped with hundreds of antennas, are an attractive way to handle the rapid growth of data traffic. As the number of users increases, the initial access and handover in contemporary networks will be flooded by user collisions. In this work, we propose a random access procedure that resolves collisions and also performs timing, channel, and power estimation by simply utilizing the large number of antennas envisioned in massive MIMO systems and the inherent timing misalignments of uplink signals during network access and handover. Numerical results are used to validate the performance of the proposed solution under different settings. It turns out that the proposed solution can detect all collisions with a probability higher than 90%, at the same time providing reliable timing and channel estimates. Moreover, numerical results demonstrate that it is robust to overloaded situations.

## I. INTRODUCTION

Massive MIMO is considered as one of the most promising solution to handle the dramatic increase of mobile data traffic in the years to come [1], [2]. The basic premise behind massive MIMO is to reap all the benefits of conventional MIMO, but on a much greater scale: a few hundred antennas are used at the base station (BS) to simultaneously serve many tens of user equipment terminals (UEs) in the same frequency-time resource using a time division duplexing protocol. This allows for coherent multi-user MIMO transmission where tens of users can be multiplexed in both the uplink (UL) and downlink (DL) of each cell. It is worth observing that, contrary to what the name "massive" suggests, massive MIMO arrays are rather compact; 160 dual-polarized antennas at 3.7 GHz fit into the form factor of a flat-screen television [3]. The benefits of massive MIMO in terms of area throughput, power consumption and energy efficiency have been extensively studied in recent years and are nowadays well understood [2], [4], [5]. On the other hand, the potential benefits of massive MIMO in the network access functionalities have received little attention [6]–[8].

Generally speaking, the network access functionalities refer to all the functions that a UE needs to go through in order to establish a radio link with the BS for data transmission and reception. To this end, the LTE standards specify a network entry procedure called random access (RA) by which uplink signals can arrive at the BS aligned in time and with approximately the same power level [9]. In its basic form, the RA function is a contention-based procedure, which essentially develops through the same steps specified by the initial ranging process of the IEEE 802.16 family of standards. Specifically, each UE trying to establish a communication link first acquires synchronization in DL and then accesses the network using the so-called random access channel (RACH), which is composed of a specified set of consecutive symbols and adjacent subcarriers. Specifically, each UE makes use of the RACH to transmit a randomly chosen code. As a consequence of the different UEs' positions within the cell, uplink signals are subject to UEs' specific propagation delays and arrive at the BS at different time instants. After detecting the selected codes and identifying how many UEs are using a given code (collision resolution), the BS proceeds extracting the corresponding timing and power information. Then, it will broadcast a response message informing the detected UEs that the random access procedure has been successfully completed and giving instructions for timing and power adjustment. The undetected UEs repeat the random access procedure until success notification. From the above discussion, it follows that code identification as well as multiuser timing and power estimation are the main tasks of the BS during the RA process.

The above procedure can be used for various purposes: initial access, handover, maintaining uplink synchronization, and scheduling request. Among the various purposes of the RA procedure and differently from [6]–[8], we focus on initial access and handover in massive MIMO networks. Both have received great attention in conventional MIMO systems and different solutions are currently available [9]–[15] (and references therein). The methods illustrated in [10] performs code detection and timing recovery by correlating the received samples with time-shifted versions of a training sequence. A simple energy detector is employed in [11] whereas a timing recovery scheme specifically devised for the LTE uplink is discussed in [9]. Schemes for initial access based on subspace methods are proposed in [12] and [13]. A solution based on the generalized likelihood ratio test is proposed in [14] whereas [15] illustrates a RA algorithm that exploits a unique ranging symbol with a repetitive structure in time-domain.

All the aforementioned schemes are developed by exploiting the degrees of freedom provided by the RACH in both time and frequency domains. In this work, we propose a novel RA protocol that aims at exploiting the spatial degrees of freedom provided by massive MIMO systems as well as the inherent different time instants of reception of UEs' signals (before the data transmission begins). The proposed procedure aims

This research has been supported by the ERC Starting Grant 305123 MORE, and by the research project 5GIOTTO funded by the University of Pisa.

at resolving collisions and also at performing timing, channel and power estimation. Specifically, it operates as follows. The number of active codes is first determined by using the minimum description length (MDL) principle [16], while the signal parameters via rotational invariance technique (ESPRIT) [17] is employed for timing recovery. The estimated timing delays are used to compute the least-square (LS) estimate of the channels of all identified UEs and also to get their own power levels. The estimated channels are eventually used by the BS in the DL to transmit a response message with "success" notification so as to inform the identified UEs that their random access procedure has been successfully completed.

## II. SYSTEM DESCRIPTION

Consider a single-cell TDD massive MIMO system using OFDM as transmission technology. We denote by $N_{\text{FFT}}$ the number of subcarriers with frequency spacing $\Delta f$. The BS is equipped with $M$ antennas and $K$ denotes the number of single-antenna UEs that would like to access the network. In doing this, each UE selects an available RACH and sends a request packet to the BS in order to notify the request of network entry. We assume that each RACH consists of $Q$ consecutive OFDM symbols and $N$ adjacent subcarriers such that $Q$ and $N$ are smaller than the (normalized) coherence time and bandwidth of all UEs, respectively. We denote $\mathbf{h}_k \in \mathbb{C}^M$ the channel frequency response of UE $k$ at the BS within the RACH and assume that

$$\mathbf{h}_k \sim \mathcal{CN}(\mathbf{0}, \beta_k \mathbf{I}_M) \quad (1)$$

where $\beta_k$ accounts for the large scale channel fading or path loss of UE $k$. Each UE transmits a *randomly* chosen code of length $Q$ in parallel over $N$ subcarriers (used for the random access procedure), which is taken from an orthogonal set $\mathcal{C} = \{\mathbf{c}_1, \mathbf{c}_2, \ldots, \mathbf{c}_Q\}$, with $\mathbf{c}_i^H \mathbf{c}_i = 1 \,\forall i$.

The signal transmitted by UE $k$ arrives at the BS with a specific carrier frequency offset (CFO) $\epsilon_k$ and a normalized (with respect to the sampling period) timing misalignment $\theta_k$. Observe that during the initial access, the CFOs are mainly due to Doppler shifts and downlink synchronization errors [18]. Therefore, they are expected to be significantly smaller than $\Delta f$. In this work, we assume that frequency estimation errors in the downlink are within 2% of $\Delta f$ as specified in [14]. In such a case, the impact of CFOs on the random access process can reasonably be neglected. On the other hand, timing errors depend on the distances of UEs from the BS. Their maximum value depends on the propagation environment, and it may reasonably be approximated by the round trip propagation delay of a UE located at the cell boundary. This parameter is known and given by $\theta_{\max} = 2R/(cT_s)$, where $R$ is the cell radius and $T_s = 1/(\Delta f N_{\text{FFT}})$ the sampling period. A simple way to counteract the effects of timing errors relies on the use of sufficiently long cyclic prefixes (CPs) comprising $N_G \geq \theta_{\max} + L$ sampling intervals, with $L$ being the maximum expected delay spread. This leads to a quasi-synchronous network in which timing errors do not produce any interblock interference and only appear as phase shifts at the output of the receive discrete Fourier transform (DFT) unit [18]. Although such a solution is normally adopted during the random access procedure, the CP of data symbols should be made just greater than the channel length to minimize unnecessary overhead. It follows that accurate timing estimates must be obtained during the random access phase in order to avoid inter-block interference in the data section of the frame.

## III. RANDOM ACCESS PROCEDURE

Under the above assumptions, the DFT output $\mathbf{y}_m(n) \in \mathbb{C}^Q$ at antenna $m$ over subcarrier $n$ during the $Q$ OFDM symbols takes the form:

$$\mathbf{y}_m(n) = \sum_{k=1}^{K} \sqrt{\rho_k} h_{km} e^{-j\frac{2\pi}{N_{\text{FFT}}}(n-1)\theta_k} \mathbf{c}_{u_k} + \mathbf{w}_m(n) \quad (2)$$

where $\mathbf{w}_m(n) \sim \mathcal{CN}(\mathbf{0}_Q, \sigma^2 \mathbf{I}_Q)$ accounts for thermal noise, $\rho_k > 0$ is the power level of the signal transmitted by UE $k$, and $u_k \in \{1, 2, \ldots, Q\}$ denotes the index of the code selected by UE $k$ for the random access. In writing the above expression, without loss of generality we have assumed that the first subcarrier of the RACH has index 0. The total power received from UE $k$ in the random access slot is thus given by

$$p_k = \rho_k \sum_{m=1}^{M} |h_{km}|^2. \quad (3)$$

Since the number of available codes is limited to $Q$, collisions may occur between different UEs which are trying to access the network. These collisions need to be detected and resolved before any UE can establish a radio link with the BS for data exchanges. In the next sections, we show how the vectors $\{\mathbf{y}_m(n)\}$ can be exploited to detect the active codes $\{\mathbf{c}_{u_k}; k = 1, \ldots, K\}$ and resolve possible collisions by exploiting the large number $M$ of antennas and the timing misalignments $\{\theta_k; k = 1, \ldots, K\}$ of UEs' signals.

*A. Code detection*

The first problem is that of determining the different code sequences which are being used, and the total number $K$ of users which are transmitting over the RACH. Since at this stage the BS has no knowledge of which codes have been selected, the complete set $\mathcal{C}$ must be considered. Therefore, for each possible code $\mathbf{c}_i$, the following quantity is computed:

$$z_{im}(n) = \mathbf{c}_i^H \mathbf{y}_m(n). \quad (4)$$

Let $\mathcal{K}_i \subseteq \{1, \ldots, K\}$ be the subset of UEs that have selected code $\mathbf{c}_i$ and denote by $K_i$ its cardinality, i.e., $K_i = |\mathcal{K}_i|$. From (2) and (4), we have

$$z_{im}(n) = \sum_{i_l \in \mathcal{K}_i} \sqrt{\rho_{i_l}} h_{i_l m} e^{-j\frac{2\pi}{N_{\text{FFT}}}(n-1)\theta_{i_l}} + \tilde{w}_{im}(n) \quad (5)$$

with $\tilde{w}_{im}(n) = \mathbf{c}_i^H \mathbf{w}_m(n)$. The samples $\{z_{im}(n); \forall n\}$ are collected into the vector $\mathbf{z}_{im} \in \mathbb{C}^N$ given by

$$\mathbf{z}_{im} = \sum_{i_l \in \mathcal{K}_i} h'_{i_l m} \mathbf{a}(\theta_{i_l}) + \tilde{\mathbf{w}}_{im} \quad (6)$$

with $\mathbf{a}(\theta_{i_l}) = [1, \ldots, e^{-j\frac{2\pi}{N_{\text{FFT}}}(N-1)\theta_{i_l}}]^T \in \mathbb{C}^N$ and $h'_{i_l m} = \sqrt{\rho_{i_l}} h_{i_l m}$. The above equation indicates that $\mathbf{z}_{im}$ has the same structure as the measurement model for a uniform linear array of passive sensors in the presence of multiple uncorrelated sources. Hence, an estimate of $K_i$ can be obtained by performing an eigendecomposition (EVD) of the correlation matrix $\mathbf{R}_{\mathbf{z}_i} = \mathbb{E}\{\mathbf{z}_{im}\mathbf{z}_{im}^H\}$. To see how this comes about, let $\lambda_{i1} \geq \lambda_{i2} \geq \cdots \geq \lambda_{iN}$ be the eigenvalues of $\mathbf{R}_{\mathbf{z}_i}$ arranged in non-increasing order. Then, from (6) it follows that

$$\lambda_{ik} = \mu_{ik} + \sigma^2 \quad\quad k = 1, \ldots, r_i \quad (7)$$
$$\lambda_{ik} = \sigma^2 \quad\quad k = r_i + 1, \ldots, N \quad (8)$$

where $r_i \leq K_i$ and $\mu_{ik} > 0$ are, respectively, the rank and the non-zero eigenvalues of the matrix $\sum_{l \in \mathcal{K}_i} \rho_l \beta_l \mathbf{a}(\theta_l) \mathbf{a}(\theta_l)^H$. Such a matrix is of full-rank, i.e. $r_i = K_i$, iff $\theta_{i_l} \neq \theta_{i_k}$ for $l \neq k$. Since the timing misalignments are continuous random variables, the probability that $\theta_{i_l} = \theta_{i_k}$ for $l \neq k$ is equal to zero, and hence $r_i = K_i$ with probability 1. This means that if the true correlation matrix $\mathbf{R}_{\mathbf{z}_i}$ were available all collisions could be in principle resolved provided that $K_i \leq N-1$. In practice, however, $\mathbf{R}_{\mathbf{z}_i}$ is not available at the receiver and must be replaced with some suitable estimate. One common approach is to use the sample correlation matrix

$$\hat{\mathbf{R}}_{\mathbf{z}_i} = \frac{1}{M} \sum_{m=1}^M \mathbf{z}_{im} \mathbf{z}_{im}^H \quad (9)$$

which provides an unbiased and consistent estimate of $\mathbf{R}_{\mathbf{z}_i}$ when $M$ is sufficiently large (as in massive MIMO systems). Performing the EVD of $\hat{\mathbf{R}}_{\mathbf{z}_i}$ and arranging the corresponding eigenvalues $\hat{\lambda}_{i1} \geq \ldots \geq \hat{\lambda}_{iN}$ in non-increasing order, we can find an estimate of $K_i$ through information-theoretic criteria. Two prominent solutions in this sense are based on the Akaike and MDL criteria. Here, we adopt the MDL approach which looks for the minimum of the following objective function [16]:

$$\hat{K}_i = \arg\min_k \frac{1}{2} k(2N-k) \ln M - M(N-k) \ln \hat{g}(k) \quad (10)$$

where $\hat{g}(k)$ is the ratio between the geometric and arithmetic means of $\{\hat{\lambda}_{in}; n = k+1, \ldots, N\}$.

### B. Timing estimation

The ESPRIT algorithm provides an elegant means for estimating the frequencies of multiple complex sinusoidal signals embedded in white Gaussian noise [17]. This method belongs to the class of subspace decomposition schemes and exploits the shift-invariant structure of the received signal to get estimates of the unknown parameters in closed form [19]. Compared to the widely used MUSIC estimator [20], ESPRIT exhibits similar accuracy with a remarkable reduction of complexity as it dispenses with computationally demanding peak search procedures. In the sequel, ESPRIT is applied to the model (5) to find an estimate of $\{\theta_{i_l}; l = 1, \ldots, K_i\}$ under the assumption that $K_i$ has been correctly estimated, i.e., $\hat{K}_i = K_i$. Observe that a fundamental assumption behind the ESPRIT estimator is that $\mathcal{K}_i$ should not exceed $N-1$. This means that the maximum number of UEs that can simultaneously select code $\mathbf{c}_i$ is limited by $N-1$.

We begin by arranging the eigenvectors of $\hat{\mathbf{R}}_{\mathbf{z}_i}$ associated to the $K_i$ largest eigenvalues into the matrix $\mathbf{V}_i = [\mathbf{v}_1 \mathbf{v}_2 \cdots \mathbf{v}_{K_i}] \in \mathbb{C}^{N \times K_i}$. The timing offsets $\{\theta_{i_l}; l = 1, \ldots, K_i\}$ of the $K_i$ UEs, which have selected code $\mathbf{c}_i$, are thus estimated in a decoupled fashion as

$$\hat{\theta}_{i_l} = \frac{N_{\text{FFT}}}{2\pi} \arg\{\psi_{il}\} \quad l = 1, \ldots, K_i \quad (11)$$

where $\{\psi_{i1}, \ldots, \psi_{iK_i}\}$ are the eigenvalues of

$$\overline{\mathbf{V}}_i = \left(\mathbf{V}_i^{(1)^H} \mathbf{V}_i^{(1)}\right)^{-1} \mathbf{V}_i^{(1)^H} \mathbf{V}_i^{(2)} \quad (12)$$

and the matrices $\mathbf{V}_i^{(1)}$ and $\mathbf{V}_i^{(2)}$ are obtained by collecting the first and the last $N-1$ rows of $\mathbf{V}_i$, respectively. The estimates $\hat{\theta}_{i_l}$ are then used for identifying the active UEs within $\mathcal{K}_i$. Clearly, the identifiability of the UEs associated to code $\mathbf{c}_i$ is guaranteed as long as the timing misalignments are different. In other words, two or more UEs that pick the same code $\mathbf{c}_i$ and are received by the BS with the same time delay are not distinguishable.

### C. Channel and power level estimation

Once timing estimation has been performed, the BS must proceed acquiring information about the channels and power levels of the users in $\mathcal{K}_i$. From (5), the LS estimate of $h'_{i_l m}$ is found to be

$$\hat{h}'_{i_l m} = \mathbf{e}_l^T \left(\hat{\mathbf{A}}_i^H \hat{\mathbf{A}}_i\right)^{-1} \hat{\mathbf{A}}_i^H \mathbf{z}_{im} \quad i_l \in \mathcal{K}_i \quad (13)$$

where $\mathbf{e}_l$ denotes the $l$th component of the canonical basis for $\mathbb{R}^{K_i}$, and $\hat{\mathbf{A}}_i \in \mathbb{C}^{N \times K_i}$ collects the vectors $\mathbf{a}(\hat{\theta}_{i_l})$ with $i_l \in \mathcal{K}_i$. Observe that $\hat{\mathbf{A}}_i$ is a Vandermonde matrix. Therefore, the full-rank condition, needed for the computation of the inverse $\hat{\mathbf{A}}_i^H \hat{\mathbf{A}}_i$, is met if and only if $\hat{\theta}_{i_k} \neq \hat{\theta}_{i_j} \forall k \neq j$.

Assume that the timing offsets have been perfectly estimated, i.e., $\hat{\theta}_{i_l} = \theta_{i_l} \forall l$, so that $\hat{\mathbf{A}}_i = \mathbf{A}_i$. Plugging (5) into (13) we obtain

$$\hat{h}'_{i_l m} = h'_{i_l m} + \epsilon_{i_l m} \quad (14)$$

where $\epsilon_{i_l m} \sim \mathcal{CN}(0, \sigma^2 [\mathbf{A}_i^H \mathbf{A}_i]_{l,l}^{-1})$. The above equation indicates that $\hat{h}'_{i_l m}$ is an unbiased estimate of $h'_{i_l m}$ with variance $\sigma^2 [\mathbf{A}_i^H \mathbf{A}_i]_{l,l}^{-1}$. Taking (3) into account, the form of (14) suggests the following ad-hoc estimator of the total power:

$$\hat{p}_{i_l} = \left(\sum_{m=1}^M |\hat{h}'_{i_l m}|^2 - M\sigma^2 [\hat{\mathbf{A}}_i^H \hat{\mathbf{A}}_i]_{l,l}^{-1}\right)^+ \quad i_l \in \mathcal{K}_i \quad (15)$$

where $(x)^+$ denotes the maximum between $0$ and $x$. The estimated channels $\{\hat{h}'_{i_l m}\}$ are eventually used by the BS in the DL to transmit a response message with a "success" notification to the identified UEs in $\mathcal{K}_i$. This message informs them that their random access procedure has been successfully completed, and contains instructions for timing and power adjustment to establish a data communication link over a set of specified subcarriers. Notice that $\hat{p}_{i_l}$ may occasionally be null for some $i_l \in \mathcal{K}_i$. This could happen, for example, when the SNR of a given user is too low or when the number of users in $\mathcal{K}_i$ has been estimated by the MDL algorithm erroneously. In this case, the BS recognizes the anomaly, and does not transmit any message to the UEs for which $\hat{p}_{i_l} = 0$.

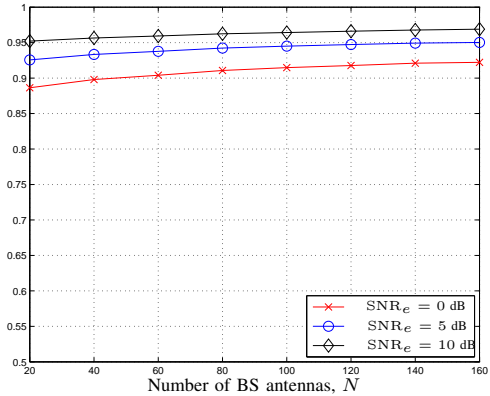

Fig. 1. Probability of estimating correctly all the UEs in each set $\mathcal{K}_i$ as a function of $M$ when $K = 15$ and $\text{SNR}_e$ is 0, 5 or 10 dB.

## IV. NUMERICAL RESULTS

Monte-Carlo simulations have been used to assess the performance of the proposed random access algorithm in terms of code detection capability as well as timing, channel, and power estimation accuracy. The results are obtained for 1000 different channel realizations and UE distributions. We assume that the UEs are uniformly distributed in a circular cell with radius normalized to one and minimum distance 0.1. The path loss function $\beta_k$ is obtained as $\beta_k = d_k^{-\kappa}$ where $d_k$ is the distance of UE $k$ from the BS and $\kappa = 3.7$ is the path loss exponent. The DFT size is $N_{\text{FFT}} = 512$ and the random access slot is composed of $Q = 8$ OFDM symbols and $N = 16$ adjacent subcarriers. Unless otherwise specified, the number of UEs that simultaneously access the network is $K = 15$, corresponding to a fully-loaded system. Each UE randomly selects a code from a Walsh-Hadamard codebook of dimension $Q = 8$. The timing misalignments $\{\theta_k\}$ are randomly taken in the interval $[0, \theta_{\max}]$ with $\theta_{\max} = 256$.

We begin by investigating the performance of the proposed algorithm in terms of probability of identifying correctly the UEs that are trying to get access to the network. Fig. 1 plots the performance of the MDL algorithm as a function of $M$ when the SNR at the cell edge, $\text{SNR}_e$, is 0, 5, or 10 dB. As seen, the MDL performs very well as the number of UEs is correctly estimated with a probability higher than 90% for all values of $\text{SNR}_e$ when $M > 40$. As expected, adding more antennas improves the system performance, but at a slow pace. Essentially, the curves in Fig. 1 show the probability that *all* the collisions within each set $\mathcal{K}_i$ are *correctly resolved* by MDL when the true correlation matrix $\mathbf{R}_{\mathbf{z}_i}$ is not available and its estimate $\hat{\mathbf{R}}_{\mathbf{z}_i}$ is used instead. In this regard, it is interesting to compare the results of Fig. 1 with the probability of collision in conventional systems in which collision-avoidance entirely rests on the choice of different code sequences (see for example [10], [14]). Assuming that $QN$ codes are available (recall that $QN$ is the total number of resources available for our system in the time-frequency domain), the probability that no collision occurs is given by:

$$p_{nc} = \frac{(QN)!}{(QN-K)!}(QN)^{-K}. \qquad (16)$$

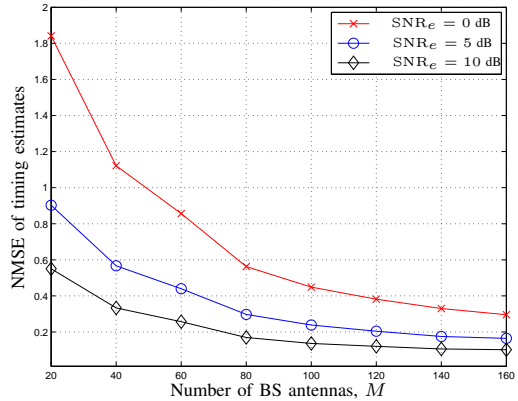

(a) NMSE of timing estimates

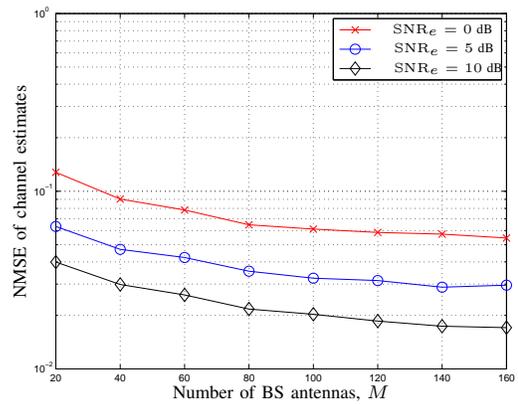

(b) NMSE of channel estimates

Fig. 2. NMSE of timing and channel estimates as a function of $M$ when $K = 15$ and $\text{SNR}_e$ is 0, 5 or 10 dB.

For example, with $QN = 128$ and $K = 15$ we have $p_{nc} = 0.426$. Contrasting this value with the results in Fig. 1, it is seen that, with the proposed procedure, the probability of identifying correctly all the UEs trying to access the network is more than double that of a conventional system. Notice that for the conventional system it has been assumed that two users with different codes can be distinguished *independently of their relative power levels* (and hence independently of the SNRs). On the other hand, the results in Fig. 1 are more realistic since they take into account the power levels of the different UEs trying to access the network.

Fig. 2 illustrates the normalized mean-square-error (NMSE) of the timing and channel estimates versus $M$ for $\text{SNR}_e = 0, 5,$ and 10 dB. For channel estimation, the normalized mean-square-error is defined as $\mathrm{NMSE}_h = \mathbb{E}\{\sum_{m=1}^{M} |h'_m - \hat{h}'_m|^2 / (\sum_{m=1}^{M} |h'_m|^2)\}$ where the statistical expectation is taken with respect to the thermal noise, the channel realizations, and the UE distribution. The results of Fig. 2(a) shows that the NMSE of the timing estimates decreases fast as $M$ grows and it is smaller than one sampling interval already for $M > 40$. From Fig. 2(b), it is seen that the NMSE of channel estimates takes small values. Therefore, the BS can reliably make use of the estimated channels in the subsequent DL phase to transmit a response message to the identified UEs.

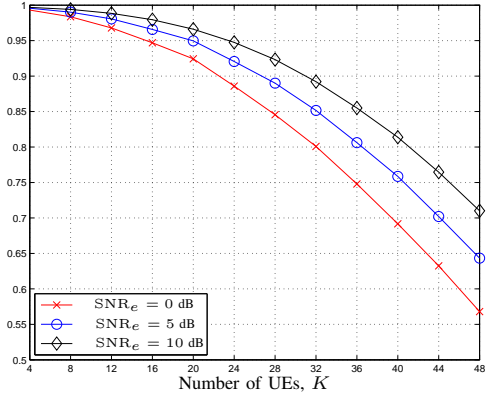

Fig. 3. Probability of estimating correctly the number of UEs in each set $\mathcal{K}_i$ as a function of $K$ when $M/K = 8$ and $\mathrm{SNR}_e$ is $0, 5$ or $10$ dB.

The results of Fig. 3 investigate the performance of the procedure as a function of $K$ when the ratio $M/K = 8$ as envisioned in massive MIMO systems. As expected, increasing $K$ deteriorates the performance of MDL since the number of collisions increases. However, a probability of correct estimation of 80% is guaranteed for $K$ up to 32.

## V. Conclusions

We proposed a random access algorithm for initial access and handover in the uplink of massive MIMO systems. By exploiting the spatial degrees of freedom provided by massive MIMO systems as well as the inherent different time instants of reception of uplink signals, the proposed solution resolved all the collisions with high probability and, at the same time, performed timing, channel and power estimation with high accuracy. Numerical results demonstrated the robustness of the proposed solution to overloaded situations.